\documentstyle[12pt]{article}
\input epsf
\newcommand{\postscript}[2] 
{\setlength{\epsfxsize}{#2\hsize}  
\centerline{\epsfbox{#1}}}
\newcommand{\nc}{\newcommand}
\nc{\bg}{B. Grz${{{\rm a}_{}}_{}}_{\hskip -0.18cm\varsigma}$dkowski}
\nc{\lsp}{\;\;\;\;\;\;\;\;}
\nc{\beq}{\begin{equation}}   \nc{\eeq}{\end{equation}}
\nc{\bea}{\begin{eqnarray}}   \nc{\eea}{\end{eqnarray}}
\nc{\baa}{\begin{array}}      \nc{\eaa}{\end{array}}
\nc{\bit}{\begin{itemize}}    \nc{\eit}{\end{itemize}}
\nc{\ben}{\begin{enumerate}}  \nc{\een}{\end{enumerate}}
\nc{\bce}{\begin{center}}     \nc{\ece}{\end{center}}
\nc{\non}{\nonumber}
\nc{\barx}{\bar{x}}
\nc{\pbarn}{\rm pb}
\nc{\fmbarn}{\rm fb}
\nc{\re}{\hbox {Re}}
\nc{\mev}{\hbox {MeV}} \nc{\gev}{\;\hbox {GeV}}
\def\gesim{\lower0.5ex\hbox{$\:\buildrel >\over\sim\:$}} 
\def\lesim{\lower0.5ex\hbox{$\:\buildrel <\over\sim\:$}} 
\nc{\prd}[3]{{\it Phys.\ Rev.}\ {{\bf D{#1}} (#2), #3}}
\nc{\prl}[3]{{\it Phys.\ Rev.\ Lett.}\ {{\bf {#1}} (#2), #3}}
\nc{\plb}[3]{{\it Phys.\ Lett.}\ {{\bf B{#1}} (#2), #3}}
\nc{\npb}[3]{{\it Nucl.\ Phys.}\ {{\bf B{#1}} (#2), #3}}
\nc{\ptp}[3]{{\it Prog.\ Theor.\ Phys.}\ {{\bf {#1}} (#2), #3}}
\nc{\zfp}[3]{{\it Z.\ Phys.}\ {{\bf C{#1}} (#2), #3}}
\nc{\mpla}[3]{{\it Mod.\ Phys.\ Lett.}\ {{\bf A{#1}} (#2), #3}}
\nc{\rmp}[3]{{\it Rev.\ Mod.\ Phys.}\ {{\bf {#1}} (#2), #3}}
\nc{\ijmpa}[3]{{\it Int.\ J.\ of\ Mod.\ Phys.}\
               {{\bf A{#1}} (#2), #3}}
\nc{\ttbar}{t\bar{t}}         \nc{\bbbar}{b\bar{b}}
\nc{\twbdec}{t\to W^+ b}
\nc{\tbwbdec}{\bar{t}\to W^- \bar{b}}
\nc{\epem}{e^+e^-}            \nc{\eett}{\epem \to \ttbar}
\nc{\sigeett}{\sigma_{e\bar{e}\to\ttbar}}  
\nc{\tbar}{\bar{t}}           \nc{\bbar}{\bar{b}}
\nc{\mt}{m_t}                 \nc{\mts}{m_t^2}
\nc{\lp}{\ell^+}              \nc{\lm}{\ell^-}
\nc{\fp}{F_+}                 \nc{\fm}{F_-}
\nc{\gp}{G_+}                 \nc{\gm}{G_-}
\nc{\fpm}{F_{\pm}}            \nc{\gpm}{G_{\pm}}
\nc{\epsl}{\epsilon_L}        \nc{\leff}{L\epsilon_{\ell\ell}}
\nc{\cpv}{$C\!P$}

\begin{document}
\pagestyle{empty} \setlength{\footskip}{2cm}
\setlength{\oddsidemargin}{0.5cm} \setlength{\evensidemargin}{0.5cm}
\renewcommand{\thepage}{-- \arabic{page} --}
\def\mib#1{\mbox{\boldmath $#1$}}
\def\bra#1{\langle #1 |}      \def\ket#1{|#1\rangle}
\def\vev#1{\langle #1\rangle} \def\dps{\displaystyle}
   \def\thebibliography#1{\centerline{REFERENCES}
   \list{[\arabic{enumi}]}{\settowidth\labelwidth{[#1]}\leftmargin
   \labelwidth\advance\leftmargin\labelsep\usecounter{enumi}}
   \def\newblock{\hskip .11em plus .33em minus -.07em}\sloppy
   \clubpenalty4000\widowpenalty4000\sfcode`\.=1000\relax}\let
   \endthebibliography=\endlist
   \def\sec#1{\addtocounter{section}{1}
       \section*{\normalsize\bf\arabic{section}. #1}\vspace*{-0.3cm}}
\vspace*{-1cm}\noindent
\hspace*{10.8cm}IFT-18-96\\
\hspace*{10.8cm}TOKUSHIMA 96-02\\
\hspace*{10.8cm}(hep-ph/9608306)\\

\vspace*{.5cm}

\begin{center}
{\large\bf$\mib{C}\!\mib{P}$-Violating Lepton-Energy Correlation in
$\mib{e}\bar{\mib{e}}\to\mib{t}\bar{\mib{t}}$}
\end{center}

\vspace*{1.25cm}
\begin{center}
\renewcommand{\thefootnote}{*)}
{\sc Bohdan
GRZ${{{\rm A}_{}}_{}}_{\hskip -0.18cm\varsigma}
$DKOWSKI$^{\: a),\: }$}\footnote{E-mail address:
\tt bohdan.grzadkowski@fuw.edu.pl}
and
\renewcommand{\thefootnote}{**)}
{\sc Zenr\=o HIOKI$^{\: b),\: }$}\footnote{E-mail address:
\tt hioki@ias.tokushima-u.ac.jp}
\end{center}

\vspace*{1.2cm}
\centerline{\sl $a)$ Institute for Theoretical Physics,\ Warsaw 
University}
\centerline{\sl Ho\.za 69, PL-00-681 Warsaw, POLAND} 

\vskip 0.3cm
\centerline{\sl $b)$ Institute of Theoretical Physics,\ 
University of Tokushima}
\centerline{\sl Tokushima 770, JAPAN}

\vspace*{1.5cm}
\centerline{ABSTRACT}

\vspace*{0.4cm}
\baselineskip=20pt plus 0.1pt minus 0.1pt
In order to observe a signal of possible $C\!P$ violation in
top-quark couplings, we have studied energy correlation of the final
leptons in $e^+e^-\to t\bar{t} \to \ell^+\ell^-X\,/\,\ell^\pm X$ at
future linear colliders. Applying the recently-proposed optimal
method, we have compared the statistical significances of
$C\!P$-violation-parameter determination using double- and
single-lepton distributions. We have found that the
single-lepton-distribution analysis is more advantageous.
\vfill
\newpage
\renewcommand{\thefootnote}{\sharp\arabic{footnote}}
\pagestyle{plain} \setcounter{footnote}{0}
\baselineskip=21.0pt plus 0.2pt minus 0.1pt

The top quark, thanks to its huge mass, is expected to provide us a
good opportunity to study beyond-the-Standard-Model physics. Indeed,
as many authors pointed $[1\ -\ 8]$, \cpv\ violation in its
production and decay could be a useful signal for possible
non-standard interactions. This is because $(i)$ the $C\!P$ violation
in the top-quark couplings induced within the SM is far negligible
and $(ii)$ a lot of information on the top quark is to be transferred
to the secondary leptons without getting obscured by the
hadronization effects.

In a recent paper, we have investigated $C\!P$ violation in the $t
\bar{t}$-pair productions and their subsequent decays at next linear
colliders (NLC) \cite{BGZH}. We have focused there on the
single-lepton-energy distributions. In this note, we study both the
double- and single-lepton-energy distributions in the process $e^+e^-
\to t\bar{t} \to \ell^+\ell^-X\,/\,\ell^\pm X$, and we compare the
expected precision of $C\!P$-violation-parameter determination in
each case. For this purpose, we apply the recently-proposed optimal
procedure~\cite{opt-96}.

Let us briefly summarize the main points of this method first.
Suppose we have a cross section
$$
\frac{d\sigma}{d\phi}(\equiv{\mit\Sigma}(\phi))=\sum_i c_i f_i(\phi)
$$
where the $f_i(\phi)$ are known functions of the location in
final-state phase space $\phi$ and the $c_i$ are model-dependent
coefficients. The goal would be to determine  $c_i$'s. It can be done
by using appropriate weighting functions $w_i(\phi)$ such that $\int
w_i(\phi)\Sigma(\phi)d\phi=c_i$. Generally, different choices for
$w_i(\phi)$ are possible, but there is a unique choice such that the
resultant statistical error is minimized. Such functions are given by
\begin{equation}
w_i(\phi)=\sum_j X_{ij}f_j(\phi)/{\mit\Sigma}(\phi)\,,
\label{X_def}
\end{equation}
where $X_{ij}$ is the inverse matrix of $M_{ij}$ which is defined as
\begin{equation}
M_{ij}\equiv \int {f_i(\phi)f_j(\phi)\over{\mit\Sigma}(\phi)}
d\phi\,.
\label{M_def}
\end{equation}
When we take these weighting functions, the statistical uncertainty
of $c_i$ becomes
\begin{equation}
{\mit\Delta}c_i=\sqrt{X_{ii}\,\sigma_T/N}\,, \label{delc_i}
\end{equation}
where $\sigma_T\equiv\int (d\sigma/d\phi) d\phi$ and $N=L_{\rm eff}
\sigma_T$ is the total number of events, with $L_{\rm eff}$ being the
integrated luminosity times efficiency.

In our analyses, we assume that only interactions of the third
generation of quarks may be affected by beyond-the-Standard-Model
physics and that all non-standard effects in the production process
($\eett$) can be represented by the photon and $Z$-boson exchange in
the $s$-channel. The effective $\gamma t\bar{t}$ and $Z t\bar{t}$
vertices are parameterized in the following form
\begin{equation}
{\mit\Gamma}^\mu=\frac{g}{2}\bar{u}(p_t)
\biggl[\,\gamma^\mu(A_v-B_v\gamma_5)
+\frac{(p_t-p_{\bar{t}})^\mu}{2\mt}(C_v-D_v\gamma_5)\,\biggr]v(p_t),
\label{vtt}
\end{equation}
\centerline{($v=\gamma$ or $Z$)}
where $g$ is the SU(2) gauge-coupling constant. In principle, there
are also four-Fermi operators which may contribute to the process of
$t\bar{t}$ production. However, as it has been verified in Ref.
\cite{bg}, their net effect is equivalent to a modification of $A_v$
and $B_v$. Therefore, without loosing generality we may restrict
ourself to the vertex corrections only.

For the on-shell $W$, we will adopt the following parameterization of
the $tbW$ vertex:
\begin{eqnarray}
&&{\mit\Gamma}^{\mu}=-{g\over\sqrt{2}}V_{tb}\:
\bar{u}(p_b)\biggl[\,\gamma^{\mu}(f_1^L P_L +f_1^R P_R)
-{{i\sigma^{\mu\nu}k_{\nu}}\over M_W}
(f_2^L P_L +f_2^R P_R)\,\biggr]u(p_t),\ \ \ \ \\
&&\bar{\mit\Gamma}^{\mu}=-{g\over\sqrt{2}}V_{tb}^*\:
\bar{v}(p_t)\biggl[\,\gamma^{\mu}(\bar{f}_1^L P_L +\bar{f}_1^R P_R)
-{{i\sigma^{\mu\nu}k_{\nu}}\over M_W}
(\bar{f}_2^L P_L +\bar{f}_2^R P_R)\,\biggr]v(p_b),
\end{eqnarray}
where $P_{L/R}\equiv(1\mp\gamma_5)/2$, $V_{tb}$ is the $(tb)$ element of 
the Kobayashi-Maskawa matrix and $k$ is $W$'s momentum.

Using the above parameterization, applying the narrow-width
approximation for the decaying intermediate particles, and assuming
that the Standard-Model contribution dominates the $C\!P$-conserving
part, we get the following normalized double- and
single-lepton-energy distributions of the reduced lepton energy
$\stackrel{\scriptscriptstyle(-)}{x}\equiv 2E\sqrt{(1-\beta)/
(1+\beta)}/m_t$, $E$ being the energy of $\ell^\pm$ in the $e^+e^-$
c.m. system, and $\beta\equiv\sqrt{1-4m_t^2/s}$ : 

\noindent
{\bf Double distribution}

\beq
\frac{1}{\sigma}\frac{d^2\sigma}{dx\;d\barx}
=\sum_{i=1}^{3}c_i f_i(x,\barx),
\label{DD}
\eeq
where $x$ and $\barx$ are for $\ell^+$ and $\ell^-$ respectively,
$$
c_1=1,\ \ \ 
c_2=\xi,\ \ \ 
c_3=\frac{1}{2}\re(f_2^R-\bar{f}_2^L)
$$
and
\bea
&&f_1(x,\bar{x})= f(x)f(\barx)+\eta'\:g(x)g(\barx)
+\eta\:[\:f(x)g(\barx)+g(x)f(\barx)\:], \non \\ 
&&f_2(x,\bar{x})= f(x)g(\bar{x})-g(x)f(\bar{x}), \non \\ 
&&f_3(x,\bar{x})=\delta\! f(x)f(\bar{x})-f(x)\delta\! f(\bar{x})
+\eta'\:[\:\delta g(x)g(\bar{x})-g(x)\delta g(\bar{x})\:] \non \\
&&\ \ \ \ \ \ \ \ \ \ \
+\eta\:[\:\delta\! f(x)g(\bar{x})-f(x)\delta g(\bar{x})
+\delta g(x)f(\bar{x})-g(x)\delta\! f(\bar{x})\:]. \non
\eea
{\bf Single Distribution}

\beq
\frac{1}{\sigma^\pm}{\frac{d\sigma}{dx}}^{\!\pm}
=\sum_{i=1}^{3}c_i^\pm f_i(x),
\label{SD}
\eeq
where $\pm$ corresponds to $\ell^{\pm}$,
$$
c_1^\pm =1,\ \ c_2^\pm =\mp \xi,\ \
c_3^+=\re{(f_2^R)},\ \ c_3^-=\re{(\bar{f}_2^L)}
$$
and
$$
f_1(x)=f(x)+\eta\:g(x),\ \
f_2(x)=g(x),\ \
f_3(x)=\delta\!f(x)+\eta\:\delta g(x).
$$

\vskip 0.6cm \noindent
Since all the functions and parameters in these formulas are to be
found in Refs.\cite{AS,BGZH}, we only remind here the normalization
of $f(x)$, $\delta\!f(x)$, $g(x)$ and $\delta g(x)$:
\begin{equation}
\int f(x)dx=1,\ \ \ \int \delta\!f(x)dx=\int g(x)dx=\int \delta
g(x)dx=0.
\end{equation}
$\eta$, $\eta'$ and $\xi$ are numerically given at $\sqrt{s}=500$
GeV as
$$
\eta=0.2021,\ \ \eta'=1.3034,\ \ 
\xi=-1.0572\:{\rm Re}(D_\gamma)-0.1771\:{\rm Re}(D_Z)
$$
for the SM parameters $\sin^2\theta_W=0.2325$, $M_W=80.26$ GeV, $M_Z=
91.1884$~GeV, ${\mit\Gamma}_Z=2.4963$ GeV and $m_t=180$ GeV.

In Eqs.(\ref{DD},\ref{SD}), $C\!P$ is violated by non-vanishing $\xi$
and/or ${\rm Re}(f_2^R-\bar{f}_2^L)$ terms.\footnote{In the present
    note, $t$, $\bar{t}$ and $W^\pm$ are assumed to be on their mass
    shell since we are adopting the narrow-width approximation for
    them, and the contribution from the imaginary part of the $Z$
    propagator is also negligible since $s$ is much larger than
    $M_Z^2$. Therefore we do not have to consider $C\!P$-violating
    effects triggered by the interference of the propagators of those
    unstable particles with any other non-standard terms\
    \cite{pilaf}.}\ 
First, let us discuss how
to observe a combined signal of $C\!P$ violation emerging via both of
these parameters. The energy-spectrum asymmetry $a(x)$ defined as
$$
a(x)\equiv\frac{d\sigma^-/dx-d\sigma^+/dx}{d\sigma^-/dx+d\sigma^+/dx}
$$
has been found as a useful measure of \cpv\ violation via $\xi$
\cite{CKP,AS}. In Ref.\cite{BGZH} we have computed $a(x)$ for the
case where both $\xi$ and ${\rm Re}(f_2^R-\bar{f}_2^L)$ terms exist.
Practically however, measuring differential asymmetries like $a(x)$
is a challenging task since they are not integrated and therefore
expected statistics cannot be high. For this reason, we shall discuss
another observable here. 

A possible asymmetry would be for instance
\begin{equation}
A_{\ell\ell}\equiv
\frac
{\dps\int\int_{x<\bar{x}}dxd\bar{x}\frac{d^2\sigma}{\dps dxd\bar{x}}
 -\int\int_{x>\bar{x}}dxd\bar{x}\frac{d^2\sigma}{\dps dxd\bar{x}}}
{\dps\int\int_{x<\bar{x}}dxd\bar{x}\frac{d^2\sigma}{\dps dxd\bar{x}}
 +\int\int_{x>\bar{x}}dxd\bar{x}\frac{d^2\sigma}{\dps dxd\bar{x}}}\:.
\label{asymm}
\end{equation}
For our SM parameters, it becomes
\begin{eqnarray}
&&A_{\ell\ell}
=0.3638\:{\rm Re}(D_\gamma)+0.0609\:{\rm Re}(D_Z)
 +0.3089\:{\rm Re}(f^R_2-\bar{f}^L_2)  \nonumber \\
&&\phantom{A_{\ell\ell}}
=-0.3441\:\xi+0.3089\:{\rm Re}(f^R_2-\bar{f}^L_2).
\end{eqnarray}
For ${\rm Re}(D_\gamma)={\rm Re}(D_Z)={\rm Re}(f^R_2)=-{\rm Re}
(\bar{f}^L_2)=0.2$, e.g., we have
$$
A_{\ell\ell}=0.2085
$$
and its statistical error is estimated to be
$$
{\mit\Delta}A_{\ell\ell}=\sqrt{(1-A_{\ell\ell}^2)/N_{\ell\ell}}
=0.9780/\sqrt{N_{\ell\ell}}.
$$
Since $\sigma_{e\bar{e}\to t\bar{t}}=0.60$ pb for $\sqrt{s}=500$
GeV, the expected number of events is $N_{\ell\ell}=600\,
\epsilon_{\ell\ell}L B_\ell^2$, where $\epsilon_{\ell\ell}$ stands
for the $\ell^+\ell^-$ tagging efficiency ($=\epsilon_\ell^2\:$;
$\epsilon_{\ell}$ is the single-lepton-detection efficiency), $L$ is
the integrated luminosity in $\fmbarn^{-1}$ unit, and $B_\ell(\simeq
0.22)$ is the leptonic branching ratio for $t$. Consequently we
obtain the following result for the error
\beq
{\mit\Delta}A_{\ell\ell}=0.1815/\sqrt{\epsilon_{\ell\ell}L},
\eeq
and thereby we are able to compute the statistical significance of
the asymmetry observation $N_{S\!D}=|A_{\ell\ell}|/{\mit\Delta}
A_{\ell\ell}$.

In Fig.1 we present lines of constant $N_{S\!D}$ as functions of
$\re{(D_\gamma)}=\re{(D_Z)}$ and $\re{(f^R_2-\bar{f}^L_2)}$ for $L=50
\ \fmbarn^{-1}$ and $\epsilon_{\ell\ell}=0.5$ (which mean $N_{\ell
\ell}=726$). Two solid lines, dashed lines and dotted lines are
determined by
$$
|\:0.4247\:\re{(D_{\gamma,Z})}+0.3089\:\re{(f^R_2-\bar{f}^L_2)}\:|
=N_{S\!D}/\sqrt{N_{S\!D}^2+N_{\ell\ell}}
$$
for $N_{S\!D}=$1, 2 and 3 respectively. We can confirm
$A_{\ell\ell}$ to be non-zero at $1\sigma$, $2\sigma$ and $3\sigma$
level when the parameters are outside the corresponding lines. It can
be seen that we have good chances for observing the effect at future
NLC unless there is a conspiracy cancellation between those
parameters. Table \ref{NSD}\ shows the $\sqrt{s}$ dependence of
$N_{S\!D}$ for the same $\epsilon_{\ell\ell}L$.

\begin{figure}[t]
\postscript{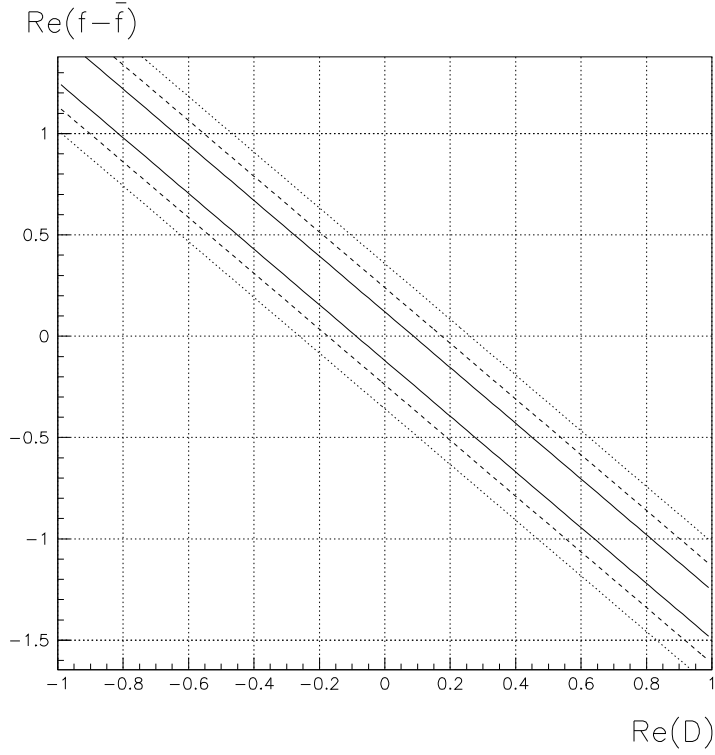}{0.75}
\caption{We can confirm the asymmetry $A_{\ell\ell}$ to be non-zero
at $1\sigma$, $2\sigma$ and $3\sigma$ level when the parameters 
$\re{(D_{\gamma,Z})}$ and $\re{(f^R_2-\bar{f}^L_2)}$ are outside the
two solid lines, dashed lines and dotted lines respectively.}
\vspace*{0.5cm}
\end{figure}
\setlength{\footskip}{2cm}
\def\ss{\scriptsize}
\def\fs{\footnotesize}
\renewcommand{\arraystretch}{1.2}
\begin{table}
\begin{center}
\vspace*{-0.3cm}
\begin{tabular}{@{\vrule width0.8pt~}c@{~\vrule width0.8pt~}cccccc
                @{~\vrule width0.8pt}} \noalign{\hrule height0.8pt}
$\sqrt{s}$ {\ss (GeV)} 
              & 500 & 600 & 700 & 800 & 900 & 1000 \\ \hline
$\sigeett$ {\ss (pb)} 
              & 0.60 & 0.44 & 0.33 & 0.25 & 0.20 & 0.16 \\
              \noalign{\hrule height0.8pt}
{\fs $P=0.1$} & 2.8 & 2.5 & 2.3 & 2.0 & 1.8 & 1.7 \\ 
              &{\fs (0.1043)}&{\fs (0.1097)}&{\fs (0.1132)} 
              &{\fs (0.1155)}&{\fs (0.1171)}&{\fs (0.1183)} \\ \hline
{\fs $P=0.2$} & 5.7 & 5.2 & 4.6 & 4.1 & 3.7 & 3.4 \\
              &{\fs (0.2085)}&{\fs (0.2195)}&{\fs (0.2263)}
              &{\fs (0.2309)}&{\fs (0.2342)}&{\fs (0.2365)} \\ \hline 
{\fs $P=0.3$} & 8.9 & 8.0 & 7.2 & 6.4 & 5.8 & 5.3 \\ 
              &{\fs (0.3127)}&{\fs (0.3292)}&{\fs (0.3395)} 
              &{\fs (0.3464)}&{\fs (0.3513)}&{\fs (0.3548)} \\ \hline
{\fs $P=0.4$} & 12.4 & 11.3 & 10.1 & 9.1 & 8.2 & 7.5 \\
              &{\fs (0.4170)}&{\fs (0.4389)}&{\fs (0.4527)} 
              &{\fs (0.4619)}&{\fs (0.4683)}&{\fs (0.4730)} \\
              \noalign{\hrule height0.8pt}
\end{tabular}
\end{center}
\caption{Energy dependence of the statistical significance $N_{S\!D}$
of $A_{\ell\ell}$ measurement for $C\!P$-violating parameters
${\rm Re}(D_\gamma)={\rm Re}(D_Z)={\rm Re}(f^R_2)=-{\rm Re}
(\bar{f}^L_2)(\equiv P)=$0.1, 0.2, 0.3 and 0.4. The numbers below
$N_{S\!D}$ (those in the parentheses) are  for the asymmetry
$A_{\ell\ell}$.}
\label{NSD}
\vspace*{0.5cm}
\end{table}       
\setlength{\footskip}{2cm}

In order to discover the mechanism of \cpv\ violation, however, it is
indispensable to separate the parameter in the top-quark production
($\xi$)\footnote{We use $\xi$ instead of $\re{(D_{\gamma,Z})}$ as a
   basic parameter when we discuss parameter measurements, since
   $\xi$ is directly related to the distributions Eqs.(\ref{DD}) and
   (\ref{SD}).}\ 
and that in the decay ($\re{(f^R_2-\bar{f}^L_2)}$). We shall apply
the optimal procedure of Ref.\cite{opt-96} to the double distribution
first. Using the functions in Eq.(\ref{DD}), we may calculate
elements of the matrix $M$ and $X$ defined in Eqs.(\ref{X_def},
\ref{M_def}):
$$
M_{11}=1,\ \ M_{12}=M_{13}=0,\ \ M_{22}=0.2070,\ \ M_{23}=-0.3368,
\ \ M_{33}=0.6049
$$
and
$$
X_{11}=1,\ \ X_{12}=X_{13}=0,\ \ X_{22}=51.3389,\ \ X_{23}=28.5825,
\ \ X_{33}=17.5662.
$$
This means the parameters are measured with errors of \footnote{Note
   that $\sigma_T$ in Eq.(\ref{delc_i}) is unity in our case since we
   are using normalized distributions.}
\beq
{\mit\Delta}\xi=7.1651/\sqrt{N_{\ell\ell}},\ \ \
{\mit\Delta}{\rm Re}(f^R_2-\bar{f}^L_2)
(=2\sqrt{X_{22}/N_{\ell\ell}})=8.3824/\sqrt{N_{\ell\ell}}\,. 
\label{D-delta}
\eeq
\setlength{\footskip}{2cm}

Next we shall consider what we can gain from the single distribution.
We have from Eq.(\ref{SD})
$$
M_{11}=1,\ \ M_{12}=M_{13}=0,\ \ M_{22}=0.0898,\ \ M_{23}=0.1499,
\ \ M_{33}=0.2699
$$
and
$$
X_{11}=1,\ \ X_{12}=X_{13}=0,\ \ X_{22}=151.9915,\ \ X_{23}=
-84.4279,\ \ X_{33}=50.6035.
$$
Therefore we get ${\mit\Delta}\xi=12.3285/\sqrt{N_{\ell}}$ and
${\mit\Delta}{\rm Re}(f^R_2)=7.1136/\sqrt{N_{\ell}}$ from the
$\ell^+$ distribution, and analogous for ${\mit\Delta}\xi$ and
${\mit\Delta}{\rm Re}(\bar{f}^L_2)$ from the $\ell^-$ distribution. 
Since these two distributions are statistically independent, we can
combine them as
\beq
{\mit\Delta}\xi=8.7176/\sqrt{N_\ell},\ \ \ 
{\mit\Delta}{\rm Re}(f^R_2-\bar{f}^L_2)=10.0601/\sqrt{N_\ell}\,. 
\label{S-delta}
\eeq

It is premature to conclude from Eqs.(\ref{D-delta}) and
(\ref{S-delta}) that we get a better precision in the analysis with
the double distribution.  As it could be observed in the numerators
in Eqs.(\ref{D-delta}, \ref{S-delta}), {\it we lose some information
when integrating the double distribution on one variable}. However,
{\it the size of the expected uncertainty depends also on the
number of events}. That is, $N_{\ell\ell}$ is suppressed by the extra
factor $\epsilon_\ell B_{\ell}$ comparing to $N_{\ell}$. This
suppression is crucial even if we could achieve $\epsilon_\ell =1$.
For $N$ pairs of $t\bar{t}$ and $\epsilon_\ell
=1$ we obtain
$$
{\mit\Delta}\xi=32.5686/\sqrt{N},\ \ \
{\mit\Delta}{\rm Re}(f^R_2-\bar{f}^L_2)=38.1018/\sqrt{N}
$$
from the double distribution, while
$$
{\mit\Delta}\xi=18.5859/\sqrt{N},\ \ \
{\mit\Delta}{\rm Re}(f^R_2-\bar{f}^L_2)=21.4484/\sqrt{N}
$$
from the single distribution.\footnote{If we take $\epsilon_\ell
    B_\ell=$0.15 as a more realistic value \cite{efficiency}, we are
    led to the same results as in \cite{BGZH}.}\ 
Therefore we may say that the single-lepton-distri-\\bution analysis
is more advantageous for measuring the parameters individually.
\setlength{\footskip}{2cm}

In summary, we have studied how to observe possible $C\!P$ violation
in $e^+ e^-\to t\bar{t}\to\ell^+\ell^-X$ and $\ell^\pm X$ at NLC. For
this purpose, $C\!P$-violating distributions of the final-lepton
energies are very useful. Using these quantities, we introduced a new
asymmetry $A_{\ell\ell}$ in Eq.(\ref{asymm}), which was shown to be
effective. Then, applying the optimal procedure \cite{opt-96}, we
computed the statistical significances of $C\!P$-violation-parameter
determination in analyses with the double- and single-lepton-energy
distributions. Taking into account the size of the leptonic branching
ratio of the top quark and its detection efficiency, we conclude
that the use of the single-lepton distribution is more
advantageous to determine each $C\!P$-violation parameter
separately.

\vspace*{0.7cm}
\centerline{ACKNOWLEDGMENTS}

\vspace*{0.3cm}
We are grateful to S. Wakaizumi for discussions
and to K. Fujii for kind correspondence on the top-quark detection
efficiency. This
work is supported in part by the Committee for Scientific Research
(Poland) under grant 2\ P03B 180 09, by Maria Sk\l odowska-Curie
Joint Found II (Poland-USA) under grant MEN/NSF-96-252, and by the
Grant-in-Aid for Scientific Research No.06640401 from the Ministry of
Education, Science, Sports and Culture (Japan).

\vspace*{0.7cm}


\begin{thebibliography}{99}
\bibitem{DV}
J.F. Donoghue and G. Valencia, \prl{58}{1987}{451}.
%
\bibitem{cprelation} 
W. Bernreuther, O. Nachtmann, P. Overmann, and T. Schr\"{o}der,
\npb{388}{1992}{53};\\
\bg\ and J.F. Gunion, \plb{287}{1992}{237}.
%
\bibitem{cptop}
F. Hoogeveen and L. Stodolski, \plb{212}{188}{505};\\
C.A. Nelson, \prd{41}{1990}{2805};\\
W. Bernreuther and O. Nachtmann, \plb{268}{1991}{424};\\
W. Bernreuther and T. Schr\"{o}der, \plb{279}{1992}{389};\\
D. Atwood and A. Soni, \prd{45}{1992}{2405};\\
G.L. Kane, G.A. Ladinsky, and C.-P. Yuan, \prd{45}{1992}{124};\\  
W. Bernreuther, J.P. Ma, and T. Schr\"{o}der, \plb{297}{1992}{318};\\
A. Brandenburg and J.P. Ma, \plb{298}{1993}{211};\\
\bg\ and W.-Y. Keung, \plb{316}{1993}{137};\\
D. Chang, W.-Y. Keung, and I. Phillips, \prd{48}{1993}{3225};\\
G.A. Ladinsky and C.-P. Yuan, \prd{49}{1994}{4415};\\
F. Cuypers and S.D. Rindani, \plb{343}{1995}{333};\\
B. Ananthanarayan and S.D. Rindani, \prd{51}{1995}{5996};\\
P. Poulose and S.D. Rindani, \plb{349}{1995}{379}, Preprints
PRL-TH-95-17 (hep-ph/9509299) and FTUV/96-36 -- IFIC/96-55
(hep-ph/9606356).
%
\bibitem{CKP}
D. Chang, W.-Y. Keung, and I. Phillips, \npb{408}{1993}{286}.
%
\bibitem{SP}
C.R. Schmidt and M.E. Peskin, \prl{69}{1992}{410}.
%
\bibitem{BG}
\bg, \plb{305}{1993}{384}.
%
\bibitem{AS}
T. Arens and L.M. Sehgal, \prd{50}{1994}{4372}.
%
\bibitem{BGZH}
\bg\ and Z. Hioki, Preprint IFT-07-96 -- TOKUSHIMA 96-01
(hep/ph-9604301 : The original version was replaced with a revised
one on August 3, 1996).
%
\bibitem{opt-96}
J.F. Gunion, \bg\ and X-G. He, Preprint UCD-96-14 -- IFT-10-96
(hep/ph-9605326).
%
\bibitem{bg}
\bg, Acta Physica Polonica {\bf B27} (1996), 921.
%
\bibitem{pilaf}
A. Pilaftsis, \zfp{47}{1990}{95};\\
M. Nowakowski and A. Pilaftsis, \mpla{6}{1991}{1933};\\
R. Cruz, \bg\ and J.F. Gunion, \plb{289}{1992}{440}.
\bibitem{efficiency}
K. Fujii, in {\it Workshop on Physics and Experiments with Linear
Colliders}, Saariselka, Finland, September 9-14, 1991;\\
P. Igo-Kemenes, in {\it Workshop on Physics and Experimentation with
Linear $e^+e^-$ Colliders},  Waikoloa, Hawaii, April 26-30, 1993.
\end{thebibliography}
\end{document}